\documentclass[proceedings, preprint]{rmaa}

\usepackage{paralist}

\usepackage{psfrag,color}

\SetYear{2014}
\SetConfTitle{Proceedings of the astrorob conference}

\title{Automated differential photometry of TAOS data: preliminary analysis} 

\author{
  D. Ricci\altaffilmark{1},
  P.-G. Sprimont\altaffilmark{2},
  C. Ayala\altaffilmark{1},
  F. G. Ram\'on-Fox\altaffilmark{1},
  R. Michel\altaffilmark{1},
  S. Navarro\altaffilmark{1},
  S.-Y. Wang\altaffilmark{3},
  Z.-W. Zhang\altaffilmark{3},
  M. J. Lehner\altaffilmark{3},
  L. Nicastro\altaffilmark{2},
  M. Reyes-Ruiz\altaffilmark{1}
}

\altaffiltext{1}{Instituto de Astronom\'ia, UNAM, Campus Ensenada,
  22860 Ensenada, B.C., M\'exico (indy@astrosen.unam.mx).}
\altaffiltext{2}{INAF, Istituto di Astrofisica Spaziale e Fisica
  Cosmica di Bologna, via P. Gobetti, 101, 40129 Bologna, Italy.}
\altaffiltext{3}{ASIAA, 11F of Astronomy-Mathematics Building,
  AS/NTU. No.1, Sec. 4, Roosevelt Rd, Taipei 10617, Taiwan, R.O.C.}

\shortauthor{Ricci et al.}
\shorttitle{RevMexAA(SC) Demo Document}

\listofauthors{D. Ricci, P-G. Sprimont, C. Ayala, F. G. Ram\'on-Fox, R. Michel,
  S. Navarro, S.-Y. Wang, Z.-W. Zhang, M. J. Lehner, L. Nicastro \& M. Reyes-Ruiz}
\indexauthor{Ricci, D.}
\indexauthor{  Sprimont, P.-G.}
\indexauthor{  Ayala,     C.  }
\indexauthor{  Ram\'on-Fox, F. G.  }
\indexauthor{  Michel,    R.  }
\indexauthor{  Navarro,   S.  }
\indexauthor{  Wang    S.-Y. } 
\indexauthor{  Zhang   Z.-W. }
\indexauthor{  Lehner M. J. }
\indexauthor{  Nicastro,  L.  }
\indexauthor{  Reyes-Ruiz, M.  }

\abstract{A preliminary data analysis of the stellar light
  curves obtained by the robotic telescopes of the TAOS project is
  presented.
  We selected a data run relative to one of the stellar fields observed
  by three of the four TAOS telescopes, and we investigate the
  common trend and the correlation between the light curves.  We
  propose two ways to remove these trends and show the preliminary results.
  A project aimed at flagging interesting behaviors, such as stellar
  variability, and to set up an automated follow-up with the San Pedro
  M\'artir Facilities is on the way.
}

\resumen{Presentamos un an\'alisis preliminar de curvas de luz
  obtenidas con datos de los telescopios rob\'oticos del proyecto
  TAOS.
  Elegimos una serie de datos en relaci\'on a uno de los campos
  estelares observados por tres de los cuatro telescopios de TAOS, e
  investigamos la tendencia com\'un y la correlaci\'on entre las
  curvas de luz. Proponemos dos maneras de borrar estas tendencias y
  presentamos los primeros resultados.
  Este es un proyecto dirigido a detectar comportamientos
  interesantes, como variabilidad estelar, que son importantes para
  establecer un seguimiento automatizado con las instalaciones del
  Observatorio de San Pedro M\'artir que est\'a actualmente en
  desarrollo.}

\addkeyword{Stars: Variables}
\addkeyword{Techniques: Photometric}
\addkeyword{Methods: Data Analysis}
\begin{document}
% Typeset article header
\maketitle

\section{General}
\label{intro}

TAOS (Taiwan American Occultation Survey) is a
project\footnote{\url{http://taos.asiaa.sinica.edu.tw/}} which makes
use of an array of four wide-field robotic $0.5 \rm m$ telescopes at the
Lulin Observatory
%\footnote{Longitude: $23^\circ28\arcmin07\arcsec N$,
%  Latitude: $120^\circ 52\arcmin 25\arcsec E$, altitude $2862\rm m$.}
(Taiwan), to monitor mainly in the ecliptic plane a total of few
thousand stars simultaneously,with high-speed photometry ($5\rm Hz$).
The main scientific goal of the project is the observation of stellar
occultations by KBO (Kuiper Belt Objects), with the aim of detecting
small (less than $1\rm km$ diameter) Transneptunian Objects
\citep{matt}.  A large amount of data is obtained every night by each
telescope, and an automatic data reduction pipeline, using the PSF
(Point Spread Function) photometry technique, has produced millions of
light curves ready for inspection \citep{kiwi}.
With the construction of TAOS II (Transneptunian Automated Occultation
Survey), the second phase of this
project\footnote{\url{http://taos2.asiaa.sinica.edu.tw/}} has started.
It is currently underway at the Observatorio Astron\'omico Nacional
%\footnote{Longitude:
%  $115^\circ 27\arcmin 49\arcsec E$, latitude: $31^\circ 02\arcmin
%  39\arcsec N$, altitude: $2830\rm m$, more than 250 clear nights per
%  year, seeing: $\approx0.6\arcsec$.} 
in San Pedro M\'artir (Baja
California, Mexico).  We are currently working to the design of the
necessary tools to take full advantage of the resulting data.
TAOS II foresees three $1.3\rm m$ telescopes that will scan several
hundreds of $3\arcmin\times 3\arcmin$ fields mainly on the ecliptic
plane, producing data that will be automatically treated to output
light curves for more than $10\ 000$ stars at a high rate ($20\rm
Hz$), for a total of over than 300TB of raw images per year.
% 
% sky: V=21.5 mag/arcsec2, R=20.7 mag/arcsec2
%
In addition to the transneptunian objects survey, which is the main
goal of the TAOS and TAOS II projects, other studies are also being
conducted with the photometric variability data: eclipsing binaries,
extrasolar planets, and variable stars \citep{var1}.
\begin{figure*}[t!]
\centering
  \includegraphics[width=0.95\textwidth]{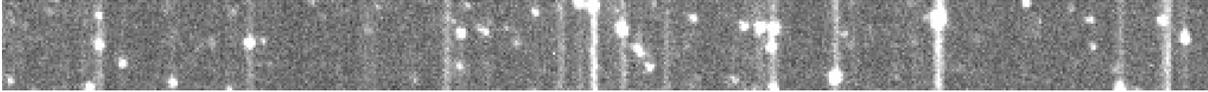}
  \caption{An example of 76-rows ``strip'' obtained reading the CCD in
    Zipper Mode.  Superimposed to the stars in the fields, the traces
    caused by the displacement of the charges while still exposing are
    clearly visible.}
  \label{zipper}
\end{figure*}
As part of an ongoing effort to design an efficient data analysis
pipeline for the large amount of data expected from the TAOS project,
we have implemented a procedure aimed at efficiently detrending the
lightcurve of the stars observed.  The same procedure can be used to
conduct ancillary scientific research using the data previously obtained.

In this paper we present the results of a preliminary analysis of TAOS
data aimed at detecting eclipsing binaries, extrasolar planets and
variable stars.  The analysis makes use of a custom software for lightcurves
detrending which combines shell scripts and C++ routines.  We also set
up a modern web based software, useful for a global inspection of the
observed fields and the lightcurve of the stars in the database
{\bf (Ricci et al., these Proceedings)}.

\section{Zipper mode}

As the time scale of the occultation events is estimated to be $\approx
200\rm ms$, a peculiar readout mode was set up to achieve high speed
imaging with the ``common'' CCD of TAOS telescopes, which has a
readout time of $2.5\rm s$.  This method is called ``Zipper
Mode''\footnote{\url{http://taos.asiaa.sinica.edu.tw/high_speed_imaging.php}},
a caractheristic that will be overrided in the second part of  the project.
The steps to perform photometry in Zipper Mode can be summarized as follows:
\begin{inparaenum}
\item expose for $105\rm ms$;
\item while the device is still exposing, move for the readout a subsection
  of 76 rows;
\item read the charges of the 76 rows (it takes a total of $95\rm ms$);
\item start next exposure of $105\rm ms$.
\end{inparaenum}
Consequently each 76-rows ``strip'' contains information concerning all the
stars of the field, as well as the traces of the stars collected
during the displacement of the charges for the readout (see Fig.~\ref{zipper}).
On each star of the ``strip'' is applied a pipeline using the PSF
(Point Spread Function) photometry technique, obtaining a light curve.
The process is repeated for each of the four telescopes.

\section{Data reduction}

\begin{figure*}[p]
\centering
  \includegraphics[width=0.85\textwidth]{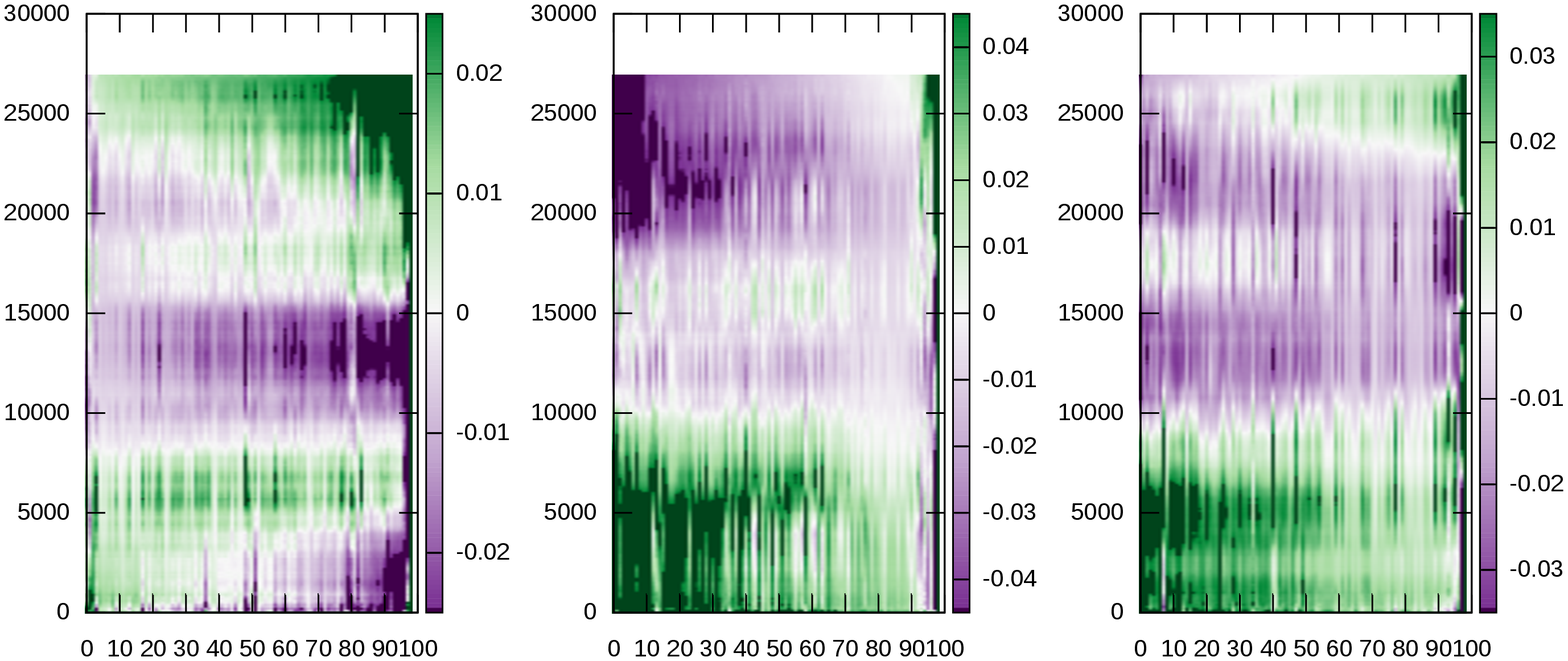} \\
  \includegraphics[width=0.85\textwidth]{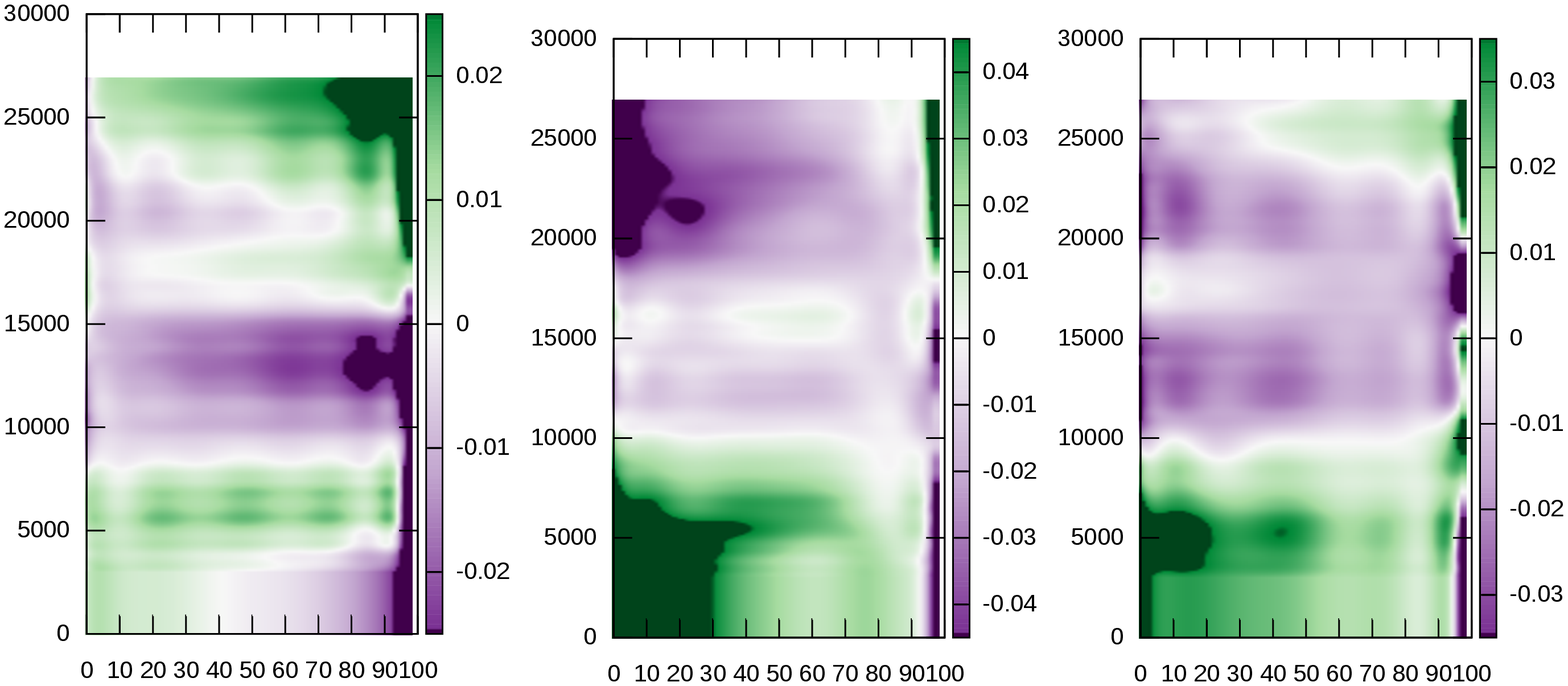} \\
  \includegraphics[width=0.85\textwidth]{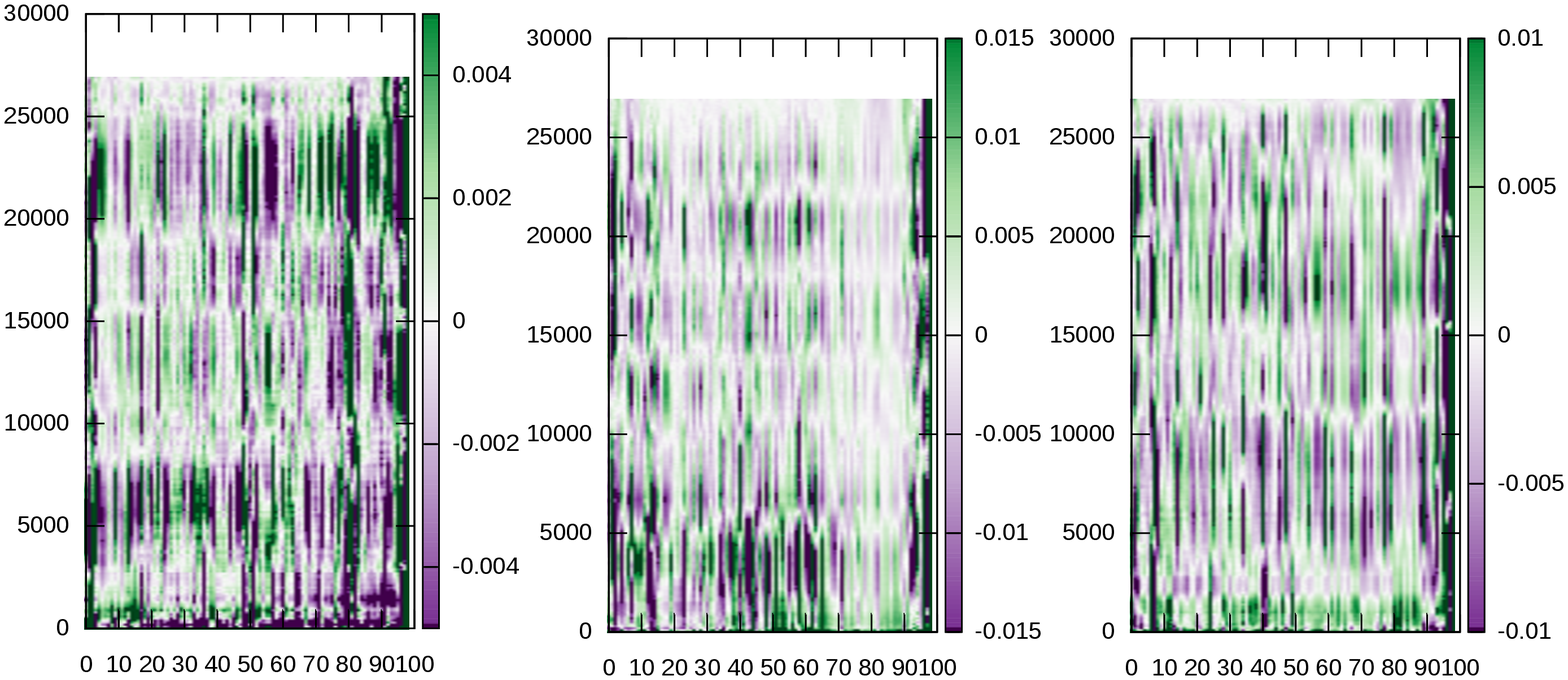}
  \caption{TOP: three plots of the light curves obtained from the
    three telescopes observing a selected TAOS field. Each of the
    100 columns represents a light curve of the 100 brightest stars in
    the field, after the subtraction of the mean magnitude value. Each
    one of the $\approx$27000 points is smoothed with a moving average
    over 3000 points (corresponding to 10 minutes). The data are
    ordered by the average of the last 300 points for visualization
    purposes. MIDDLE: two-dimensional spline model of the light
    curves. BOTTOM: the data after subtracting the model (please note
    the changes of the limits in the color scale). }
  \label{subtracted2}
\end{figure*}
The four light curves of each one of the hundreds of stars are stored
in hundreds FITS binary tables, while timing and other metadata, which
are the same for all the stars, are stored in the binary tables of a
separate FITS file.
We extracted the information about the fields observed and those
of the selected stars in two \verb|mysql| tables.
The goal of this work is to provide a reduction technique which would
allow us to effectively identify and study the following events:
\begin{inparaenum}
\item exoplanetary transits,
\item GRB afterglows,
\item supernovae,
\item variable and other transient events.
\end{inparaenum}
To perform the quick view of the data and inspect the star fields, we
built a web GUI using a web 2.0 approach (also presented in this
conference).  For our tests we chose the first run of\textbf{ one of TAOS
fields, namely the number 152, which was observed with three telescopes}, and the
100 brightest stars detected.

%\section{Data reduction}

The data are smoothed every 3000 points (5 points per seconds, i.e. 10
minutes) using a moving average, and the mean magnitude of each curve
is subtracted.  We noticed a common behaviour between curves,
different from telescope to telescope, that we ascribe to the
variations of sky conditions during the night, and a trend
``evolving'' some way from light curve to light curve. This trend does
not depend from the CCD position and it is very slightly magnitude
dependent. We decided to re-order the curves by the value of the
average of their last 300 points, for visualisation purposes only. The
result is visible in the top three plots of Fig.~\ref{subtracted2},
one for each of the three telescopes used to observe the selected
field.
We modeled the ``evolving'' trend with a two-dimensional spline using
a fast C++ routine\footnote{The routine is part of a suite whose
  components are also used in the \textit{Sadira} project
  {\bf(Sprimont et al., these Proceedings)}}, varying the number of
knots and the degree of the polynomials to obtain a consistent match.
The result can be seen in the three middle plots of
Fig.~\ref{subtracted2}.
Subtracting the model, we obtain the detrended curves with a scatter
of $4\rm mmag$, $15\rm mmag$, and $10\rm mmag$ peak-to-valley for each
telescope, respectively (see the bottom plots of
Fig.~\ref{subtracted2}).
The result is efficient, but this the goodness of the fit
depends on the curves sorting order.

\begin{figure*}[t!]
\centering
  \includegraphics[width=0.80\textwidth]{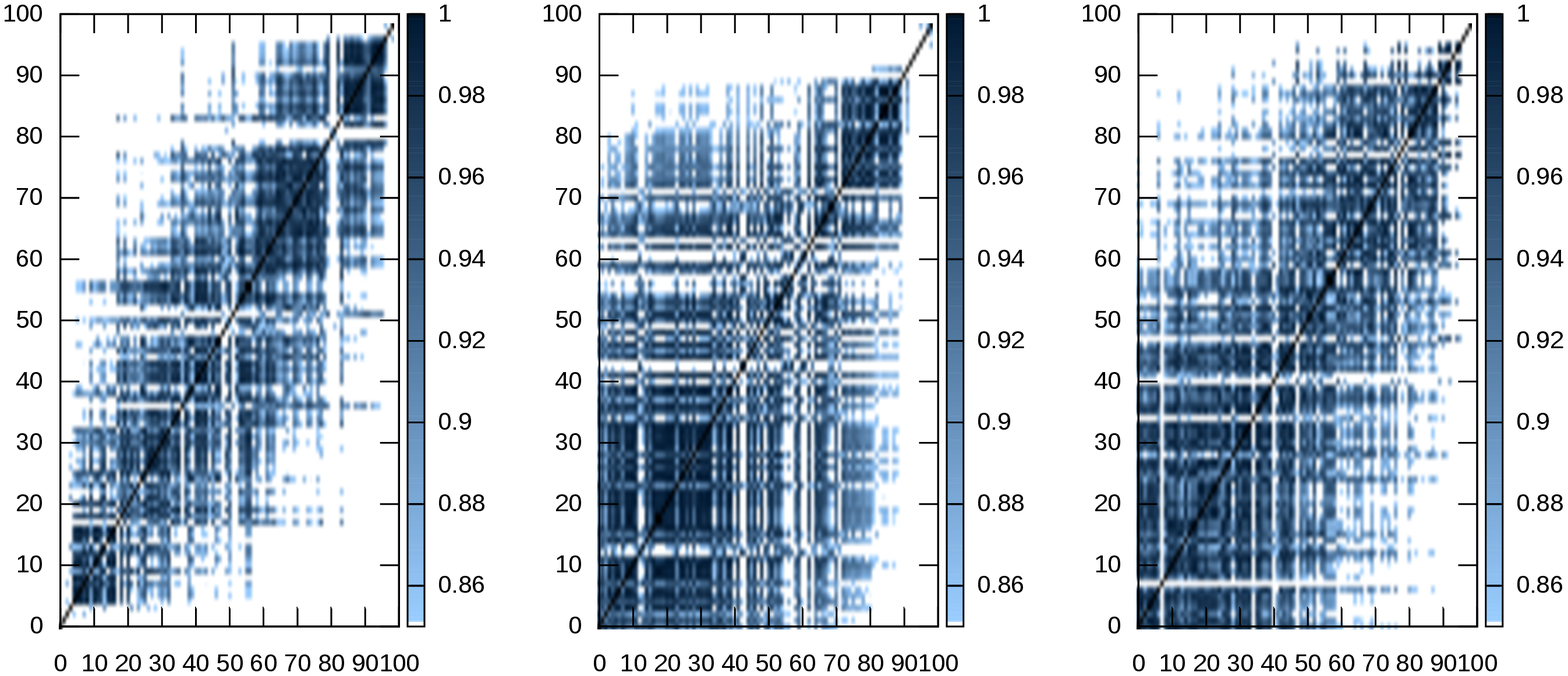}
  \caption{Correlation matrices for each of the 100 light
    curves and for the three telescopes that observed the selected TAOS
    field.  Please note that the color scale selects only 
    correlation levels greater than 85\%.}
  \label{correlation}
\end{figure*}

\begin{figure*}[t!]
\centering
  \includegraphics[width=0.85\textwidth]{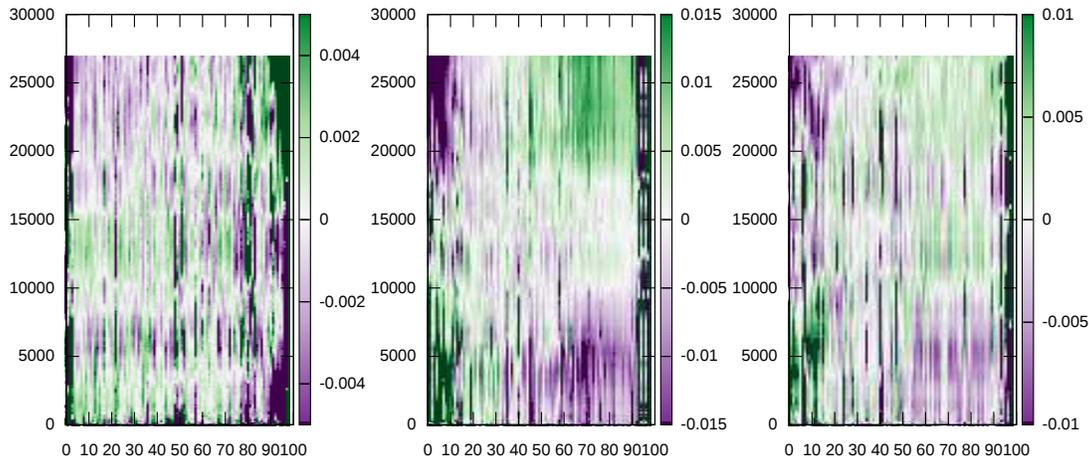}
  \caption{Light curves after subtracting an average curve obtained by
    weighting the others for their correlation factor. 
%
%    TOP: after subtracting a common behavior (mean of all the curves).
%    BOTTOM: without subtracting the common behavior.  
}
  \label{weighted}
\end{figure*}

We then took a step back and calculated the correlation factor
between each couple of curves. The three correlation matrices are shown in
Fig.~\ref{correlation}.  Moreover, we subtracted to each curve an average of
the other curves obtained by weighting each of them by their correlation
factor.  We obtain a magnitudes scatter similar to that of the previous
method (2-d spline): $4\rm mmag$, $15\rm mmag$, and $10\rm
mmag$ peak-to-valley for each of the three telescopes (see
Fig.~\ref{weighted}), but in this case the result is
sorting-independent.  Systematic trends are still present as a second
order effect, and the causes are still under investigation.

\section{Results and Follow-up}

The preliminary results of an ongoing analysis involving the light
curves of the 100 brightest star selected from a test TAOS field
exhibit similar behaviors.  In particular correlated trends in the
data set are evident.  While the causes are under
investigation, we are able to remove these effects obtaining
magnitude scatters in the range $4$--$15\rm mmag$, depending on the
considered telescope.

An automated system to flag peculiar behaviors, such as stellar
variability in the fields, is under development.
%, and the observation
%of a field containing the known variable star $\epsilon Eri$, 
It will represent the first step for the realization of the
follow-up system to track interesting events by using San Pedro
M\'artir facilities.

To this aim we are testing the multichannel photometer RATIR on the
$1.5\rm m$ telescope and the MEXMAN instrument of the $84\rm cm$
telescope by observing already known exoplanetary transits.
This way we will be able to evaluate
the capability of these instruments and the limits of the follow-up.

% We then decided to subtract as common behavior, as a first step, the
% mean of each curve. the result is shown in Fig.~\ref{noback}.
% \begin{figure*}[t!]
% \centering
%   \includegraphics[width=0.90\textwidth]{noback.eps}
%   \caption{We then tried }
%   \label{noback}
% \end{figure*}

% \begin{thebibliography}
% \bibitem[Arthur \& Hoare(2006)]{2005astro.ph.11035A} Arthur, S.~J., \&
% Hoare, M.~G.\ 2006, \apj, in press (astro-ph/0511035)
% \end{thebibliography}

\bibliography{ricci-talk}{}

\end{document}